\documentclass{iau} 
\usepackage{graphicx}

\title[Heterodyne Receivers] 
{New Laboratory Techniques using heterodyne Receivers}

\author[Nadine Wehres]   
{Nadine Wehres, 
Kirill Borisov, \\
Katharina von Schoeler, Patrick P{\"u}tz, \\
Cornelia E. Honingh, Frank Lewen  
 \and Stephan Schlemmer
 }

\affiliation{I. Physics Institute, University of Cologne, \\ Z{\"u}lpicher Strasse 77,
50937 K{\"o}ln, Germany \\ email: {\tt wehres@ph1.uni-koeln.de} }

\pubyear{2019}
\volume{350}  
\jname{Laboratory Astrophysics: from Observations to Interpretation }
\editors{A.C. Editor, B.D. Editor \& C.E. Editor, eds.}
\begin{document}

\maketitle

\begin{abstract}
Two laboratory emission spectrometers have been designed and described previously.  Here, we present a follow-up study with special focus on absolute intensity calibration of the new SURFER-spectrometer (SUbmillimeter Receiver For Emission spectroscopy of Rotational transitions), operational between 300 and 400~GHz and mostly coincident with ALMA (Atacama Large Millimeter/submillimeter Array) Band 7.  

Furthermore, we present a feasibility study to extend the detection frequencies up to 2~THz. First results have been obtained using the SOFIA (Stratospheric Observatory for IR Astronomy) upGREAT laboratory setup at the University of Cologne.  Pure rotational spectra of the complex molecule vinyl cyanide have been obtained and are used to give an estimate on the sensitivity to record ro-vibrational transitions of molecules with astrophysical importance at 2~THz.
      
\keywords{methods: laboratory, techniques: spectroscopy, molecular processes, absolute intensities}
\end{abstract}

\firstsection 
\section{Introduction}

We designed two new laboratory spectrometers with the goal to use the benefits of receiver technology for radio astronomy in a laboratory environment.  One experiment includes a room-temperature heterodyne receiver which employs state-of-the-art low-noise amplifiers operational between 70 and 110~GHz.  The second receiver uses a cryogenically cooled SIS mixer (superconducting-insulating-superconducting).  Both spectrometers have been described elsewehere (\cite{Wehres_2018a,Wehres_2018b}).  For similar efforts at lower frequencies see also the publication by \cite{Tanarro_2018}.

\begin{figure}[b]
\vspace*{-0.5 cm}
\begin{center}
\includegraphics[width=0.85\textwidth]{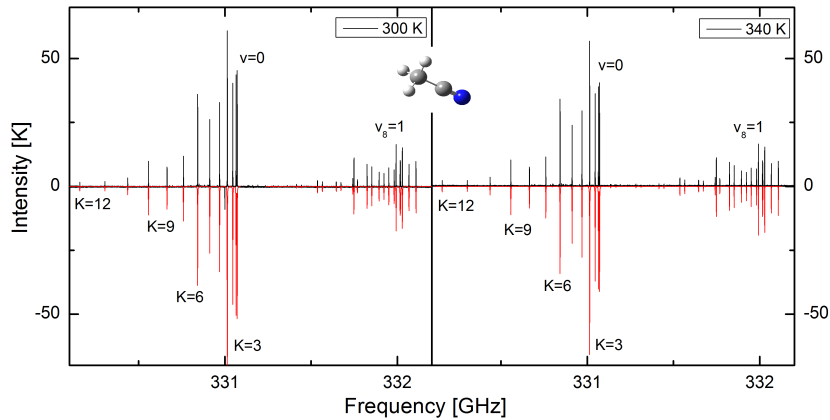} 
\caption{SURFER Emission spectrum of methyl cyanide at 300 and 340~K shown in black.  In red, a simulated and inverted spectrum using Pgopher is shown.  The intensities are well reproduced for the vibrational ground state, $\nu$=0 and the first vibrationally excited state $\nu_8$=1.  }
 \label{fig1}
\end{center}
\end{figure}

\begin{figure}[b]
\vspace*{-0.5 cm}
\begin{center}
 \includegraphics[width=0.8\textwidth]{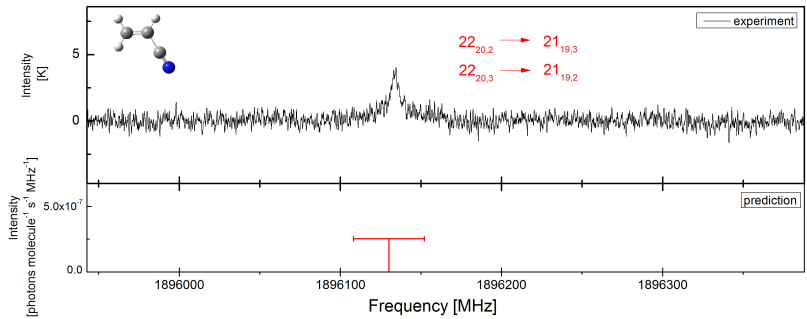} 
 \caption{Spectrum of vinyl cyanide obtained around 2~THz with the SOFIA upGREAT lab setup using 300~s integration time.  The predictions include parameters obtained from \cite{Kisiel_2012}.}
   \label{fig2}
\end{center}
\end{figure}

\begin{figure}[b]
\begin{center}
\includegraphics[width=0.40\textwidth]{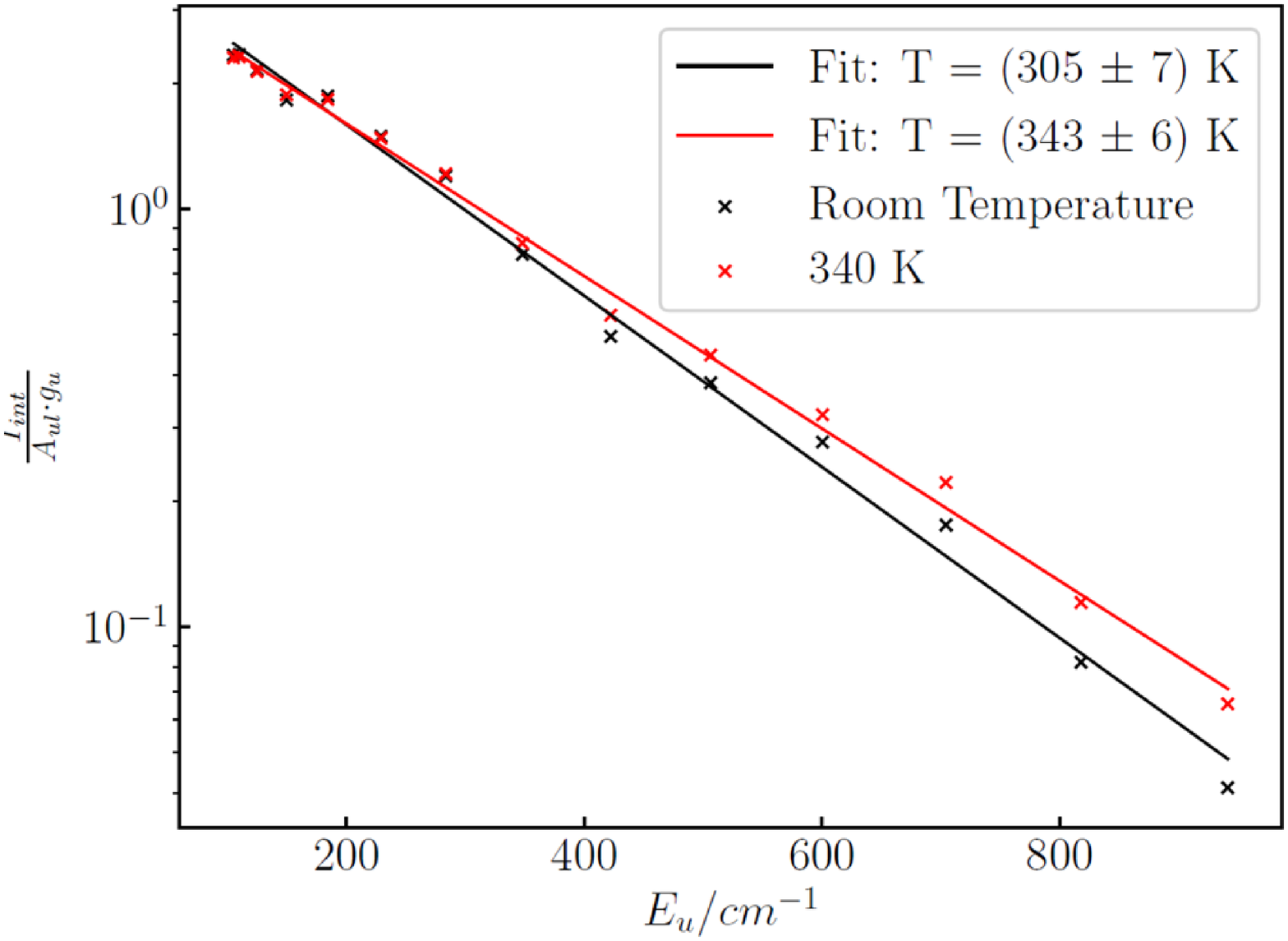} 
\caption{Boltzmann diagram:  The emission spectra of the ground vibrational state of methyl cyanide at 300~K and at 340~K have been used to determine rotational temperatures.  The discrepancy between the laboratory temperature and the calculated temperature is about 2\% .}
 \label{fig3}
\end{center}
\end{figure}

Here, we present a recent study with special focus on absolute intensities that can be obtained using the SURFER spectrometer.  This study is used to get an estimate on the expected intensities of complex molecules at higher frequencies.  
Preliminary studies using the SOFIA upGREAT laboratory setup show, that its feasible to extend the frequency coverage up to the THz regime where low-lying vibrational bands of complex molecules can be studied.  Here, we give a short outlook on the possibility to study the fingerprint-like ro-vibrational bands of complex molecules around 2~THz.

\section{Experiments}
\subsection{Emission spectroscopy with SURFER at 350~GHz}
SURFER is operational between 300 and 400~GHz and has been described elsewhere (\cite{Wehres_2018b}).  
In short: the radio frequency (RF) signal from the molecule of interest (here: methyl cyanide CH$_3$CN) is combined with a fixed frequency RF signal of a local oscillator (LO) in the SIS mixer.  The intermediate frequency (IF: 5.1--7.6~GHz) is amplified, mixed down to DC--2.5~GHz, and processed using a Fast Fourier Transform digitizer as the backend spectrometer.  This technology is commonly used at radio telescopes and their high sensitivity and bandwidth are also interesting in a laboratory environment. 

Here, the pressure in the gascell is kept stable at 1$\times$10$^{-2}$~mbar.  Integration times of 100~s are used, resulting in an absolute noise equivalent temperature of 0.1~K.  Overall, for each intensity calibrated spectrum we record three acquisitions:  First, the molecular emission in front of a cold background.  Second, the molecular emission in front of a hot background.  Third, the empty gascell using a cold background.  Microwave absorber foam is used as a calibrated blackbody, once at liquid nitrogen temperatures and once at room-temperature.  Two spectra, at 300~K and at 340~K, are presented in Fig.~\ref{fig1} covering one spectral setting of 2.5~GHz.

\subsection{Emission spectroscopy with the SOFIA upGREAT lab setup at 2~THz}

The SOFIA upGREAT lab setup in Cologne has been used to demonstrate the feasibility of heterodyne detection at higher frequencies, i.e. at 2~THz.  The setup consists of a synthesizer and VDI-amplifier-multiplier chain S209 which upconverts the frequency output to around 2~THz.  This LO signal is then combined with the molecular emission signal in the (4~K) cooled HEB (hot electron bolometer) mixer.  The IF, here 0.5--5~GHz, is further amplified and processed using a digital Fast Fourier Transform spectrometer.  Overall the HEB-mixer is a replicate of the SOFIA upGREAT receiver with very similar specifications (\cite{Risacher_2016,Putz_2015}).  The molecule (vinyl cyanide C$_2$H$_3$CN) is injected in a stainless steel gascell of roughly 30~cm length and 5~cm in diameter.  Pressures in the gascell are kept at 1~$\times$1$^{-1}$mbar.  A total integration time of 5~minutes is used to acquire the spectrum shown in Fig.~\ref{fig2}.

\section{Results:  Intensity Accuracy}
\subsection{Methyl cyanide using SURFER}
The rotational emission spectrum of methyl cyanide is shown in Fig.~\ref{fig1}.  On the left handside the spectrum at room-temperature is shown; on the right handside at elevated temperature, obtained by heating the gascell to 340~K.  
Overall, at 331~GHz and 332~GHz the pure rotational transitions of the vibrational ground state and first vibrationally excited state ($\nu_8$=1) are identified.  While the black spectrum shows the experimental spectrum, the data plotted in red show a simulation obtained with Phopher (\cite{Western_2017}) using temperatures of 300 and 340~K, respectively.  The molecular parameters were adopted from the Cologne database for molecular spectroscopy (\cite{Muller_2005}).  The intensities in the simulation are normalized to the intensities in the emission experiment.

The rotational transitions from the ground state at 300~K and at 340~K have then been used to determine the rotational temperature of methyl cyanide.  Fig.~\ref{fig3} shows a Boltzmann fit for the room-temperature spectrum and for the spectrum at elevated temperatures.  Overall, the deviation of the rotational temperature at room-temperature is about 2\%.  The deviation of the signal intensities at elevated temperatures is accurate to about 1\%.  This demonstrates that the emission signals of molecules are calibrated with high accuracy and therefore, emission spectra can be used to determine the temperature of a gas or to identify transitions where experimental intensities differ from theoretical predictions indicating well known phenomena such as so-called intensity borrowing.

\subsection{Vinyl cyanide using the SOFIA upGREAT lab setup}
Part of the rotational spectrum of vinyl cyanide (C$_2$H$_3$CN) is presented in Fig.~\ref{fig2}.  The main purpose of this proof-of-concept study is to obtain an estimate on the expected integration time for low-lying vibrational bands of complex molecules around 2~THz, i.e. the detection band of the SOFIA upGREAT receiver.  Thus, we calculated vibrational frequencies and intensities using Gaussian09 (\cite{Frisch_2009}) for some cyclic molecules and specific PAHs with low-lying vibrational bands around 2~THz.  

Cyclobutanone (C$_4$H$_6$0) shows a vibrational band ($\nu_{2\leftarrow3}$) at 65~cm$^{-1}$ (1.95~THz) (\cite{Borgers_1966}).  Calculated IR intensities on a B3LYP/aug-cc-pVDZ level are on the order of 13.4~km/mole.  This can be converted into a dipole moment of 0.28~Debye (\cite{Bernath_2005}), giving emission intensities of around 1~$\times~10^{-7}$ photons molecule$^{-1}$ s$^{-1}$ MHz$^{-1}$.  This is very comparable to the emission spectrum of vinyl cyanide and its predicted intensities ($\mu_b$=0.9~Debye) shown in Fig.~\ref{fig2}.  Here, a factor of three less in intensity results in 9 times longer integration times, i.e. 45 minutes for a spectrum of cyclobutanone.

Circumpyrene (C$_{42}$H$_{16}$) shows a vibrational band at 74~cm$^{-1}$ (2.2~THz) and calculated IR intensities are on the order of 2.8~km/mole (\cite{Malloci_2007}) or 0.13~Debye, resulting in integration times of 200 minutes.

\section{Conclusions and Outlook}
The far IR regime is characterized by the fingerprint-like ro-vibrational transitions of large molecules, such as PAHs.  Since chemically related compounds, such as other PAHs, have very similar spectral signatures, these molecules cannot be discriminated in low resolution IR studies.  Moreover, the high symmetry of most PAHs prohibits a pure rotational spectrum and therefore prevents targeting these molecules with radio telescopes.  However, low-lying vibrational bands can fall into the THz region where the rotational structure of the vibrational bands can be resolved leading to the typical $P$, $Q$ and $R$ structure. The rotationally resolved sub-structure of these bands can be used to unambiguously identify specific PAHs in astronomical observations.  
  
Overall, our preliminary study on absolute intensity calibration and thus on the expected intensities for low-lying ro-vibrational transitions of PAHs are a promising way to test the PAH hypothesis and to potentially identify PAHs using the ro-vibrational fingerprint in the far IR regime.

\section*{Acknowledgement}
Developments in heterodyne emission spectroscopy are carried out within the Collaborative Research Centre 956, sub-project [B4], funded by the Deutsche Forschungsgemeinschaft (DFG) – project ID 184018867.


\vspace*{-0.5 cm}
\begin{thebibliography}{}
\bibitem[Bernath, P.F. (2005)]{Bernath_2005} Bernath, P.F. \ 2005 Spectra of Atoms and Molecules, Second Edition, pp.~275-278 

\bibitem[Borgers, \& Strauss(1966)]{Borgers_1966} Borgers, T.~R., \& Strauss, H.~L.\ 1966, 
\textit{J. Chem. Phys.}, 45, 947

\bibitem[Frisch et al.(2009)]{Frisch_2009} Frisch, M.J., et al. 2009, Gaussian 09, Gaussian, Inc., Wallingford CT

\bibitem[Kisiel et al.(2012)]{Kisiel_2012} Kisiel, Z., Pszcz{\'o}{\l}kowski, L., Drouin, B.~J., et al.\ 2012,
\textit{J. Mol. Spectrosc.}, 280, 134

\bibitem[Malloci et al.(2007)]{Malloci_2007} Malloci, G., Joblin, C., \& Mulas, G.\ 2007, 
\textit{Chem. Phys.}, 332, 353

\bibitem[M{\"u}ller et al.(2005)] {Muller_2005} M{\"u}ller, H.~S.~P., Schl{\"o}der, F., Stutzki, J., et al.\ 2005, 
\textit{J. Mol. Struct.}, 742, 215

\bibitem[P{\"u}tz et al. (2015)]{Putz_2015} P{\"u}tz, P., B{\"u}chel, D., Jacobs, K., Schultz, M., and Honingh, C.E. \ 2015, 
\textit{Proceedings of the twenty-sixth international symposium on space Terahertz technology}, M2-3 


\bibitem[Risacher et al.(2016)]{Risacher_2016} Risacher, C., Gusten, R., Stutzki, J., et al.\ 2016, 
\textit{IEEE Trans. THz Sci. Technol.}, 6, 199

\bibitem[Tanarro et al.(2018)]{Tanarro_2018} Tanarro, I., Alem{\'a}n, B., de Vicente, P., et al.\ 2018, \textit{A\&A}, 609, A15

\bibitem[Wehres \etal\ (2018a)]{Wehres_2018a} 
{Wehres, N., Heyne, B., Lewen, F., Hermanns, M., Schmidt, B., Endres, C., Graf, U.U., Higgins, D.R., Schlemmer, S.} 2018a, 
\textit{Astrochemistry VII:  Through the Cosmos from Galaxies to Planets}, 
Proc. IAUS 332  

\bibitem[Wehres \etal\ (2018b)]{Wehres_2018b} 
{Wehres, N., Ma{\ss}en, J., Borisov, K., Schmidt, B., Lewen, F., Graf, U.U., Honingh, C.E., Higgins, D.R., Schlemmer, S.} 2018b, 
\textit{Phys. Chem. Chem. Phys.}, 20, 5530 

\bibitem[Western(2017)]{Western_2017} Western, C.~M.\ 2017, 
\textit{J. Quant. Spec. Rad. Trans.}, 186, 221




\end{thebibliography}
\end{document}